\documentclass[]{spie}  %>>> use for US letter paper
\usepackage{graphicx}% Include figure files
\usepackage{dcolumn}% Align table columns on decimal point
\usepackage{color}

\usepackage{amsmath}
\usepackage{amssymb}

\usepackage[tight,normalsize,sf,SF]{subfigure}
\title{Scalable background-limited polarization-sensitive detectors for mm-wave applications}

\author{Karwan Rostem\supit{a,b}, Aamir Ali\supit{a}, John W. Appel\supit{a}, Charles L. Bennett\supit{a}, David T. Chuss\supit{b}, Felipe A. Colazo\supit{b}, Erik Crowe\supit{b}, Kevin L. Denis\supit{b}, Tom Essinger-Hileman\supit{a}, Tobias A. Marriage\supit{a}, Samuel H. Moseley\supit{b}, Thomas R. Stevenson\supit{b}, Deborah W. Towner\supit{b}, Kongpop U-Yen\supit{b}, Edward J. Wollack\supit{b}, 
	\skiplinehalf
	\supit{a}The Johns Hopkins University, Department of Physics and Astronomy, 3400 North Charles Street, Baltimore, MD 21218. \\
	\supit{b}Goddard Space Flight Center, 8800 Greenbelt Road, MD 20771; \\
}

\authorinfo{Further author information: (Send correspondence to K. Rostem)\\ E-mail: karwan.rostem@nasa.gov, Telephone: +1 301 286 0308}  

\begin{document} 
\maketitle 
%%%%%%%%%%%%%%%%%%%%%%%%%%%%%%%%%%%%%%%%%%%%%%%%%%%%%%%%%%%%% 
%>>>> uncomment following for page numbers
% \pagestyle{plain}    
%>>>> uncomment following to start page numbering at 301 
%\setcounter{page}{301} 
 
%%%%%%%%%%%%%%%%%%%%%%%%%%%%%%%%%%%%%%%%%%%%%%%%%%%%%%%%%%%%% 
\begin{abstract}
We report on the status and development of polarization-sensitive detectors for millimeter-wave applications. The detectors are fabricated on single-crystal silicon, which functions as a low-loss dielectric substrate for the microwave circuitry as well as the supporting membrane for the Transition-Edge Sensor (TES) bolometers. The orthomode transducer (OMT) is realized as a symmetric structure and on-chip filters are employed to define the detection bandwidth. A hybridized integrated enclosure reduces the high-frequency THz mode set that can couple to the TES bolometers. An implementation of the detector architecture at Q-band achieves 90\% efficiency in each polarization. The design is scalable in both frequency coverage, 30-300 GHz, and in number of detectors with uniform characteristics. Hence, the detectors are desirable for ground-based or space-borne instruments that require large arrays of efficient background-limited cryogenic detectors. 

\end{abstract}

%>>>> Include a list of keywords after the abstract 

\section{\label{sec:intro}INTRODUCTION}

Millimeter-wave observations are extremely valuable for the study of cosmic evolution. For instance, the Sunyaev-Zeldovich effect observed below the arcminute scale is a powerful tool for probing galaxy cluster dynamics and structure growth as a function of redshift. Above the degree scale, the polarization of the Cosmic Microwave Background (CMB) is potentially the only known source accessible to measurements that is sensitive to the Big Bang initial conditions. With this perspective, the recent BICEP2 results present an exciting future for inflation cosmology~\cite{bicep2}, but also highlight the importance of distinguishing the cosmological signal from sources such as polarized galactic foregrounds~\cite{mortonson,flauger}. Hence, the relatively low source surface brightness at millimeter wavelengths poses strict requirements on an instrument and detection system, whether in a sub-orbital or space-based platform. Specifically, an instrument capable of multi-wavelength operation with control over potential systematics (e.g. symmetric beam with control over side-lobes, stray light mitigation), and high stability (low 1/$f$) is required. Ideally, the detectors in the focal-plane should be background-limited, efficient, and scalable with highly uniform characteristics across the focal-plane array. 
 
In this paper, we report on progress of Q-band ($f_c$ = 26.4 GHz; operational bandwidth 29-50 GHz) and W-band ($f_c$ = 59.1 GHz; operational bandwidth 65-112 GHz) feedhorn-coupled polarization-sensitive detectors designed for CMB observations. The detectors are fabricated on single-crystal silicon and designed with a high degree of symmetry at the circuit level that maximizes the coupling of the sky signal to the Transition-Edge Sensor (TES) bolometers for each polarization channel. The detectors are designed for the ground-based Cosmology Large Angular Scale Surveyor (CLASS) telescope~\cite{tom}. CLASS will target detection of the low-multipole reionization bump theoretically predicted in the B-mode angular power spectrum, where the signal from gravitational lensing is sub-dominant to that of inflation. To remove galactic foregrounds, CLASS will observe at four frequency bands with center frequencies of 38, 93, 148, and 217 GHz. Further details on the CLASS instrument~\cite{tom} and the 38 GHz focal plane development~\cite{john} can be found in these proceedings.

We emphasize that the scalability, uniformity, efficiency, wide bandwidth, and noise properties of the detectors described here are highly desirable for space- and balloon-borne instruments. The architecture has been demonstrated at Q-band, is currently being developed for two W-band receivers each with 259 dual-polarization detectors~\cite{tom}, and can be implemented at up to 300 GHz. 

%%%%%%%%%%%%%%%%
\section{Progress on Q-band Detectors}
\label{sec:40GHz}

\begin{figure}[!t]
\begin{center}
\subfigure[]{\includegraphics[width=8cm]{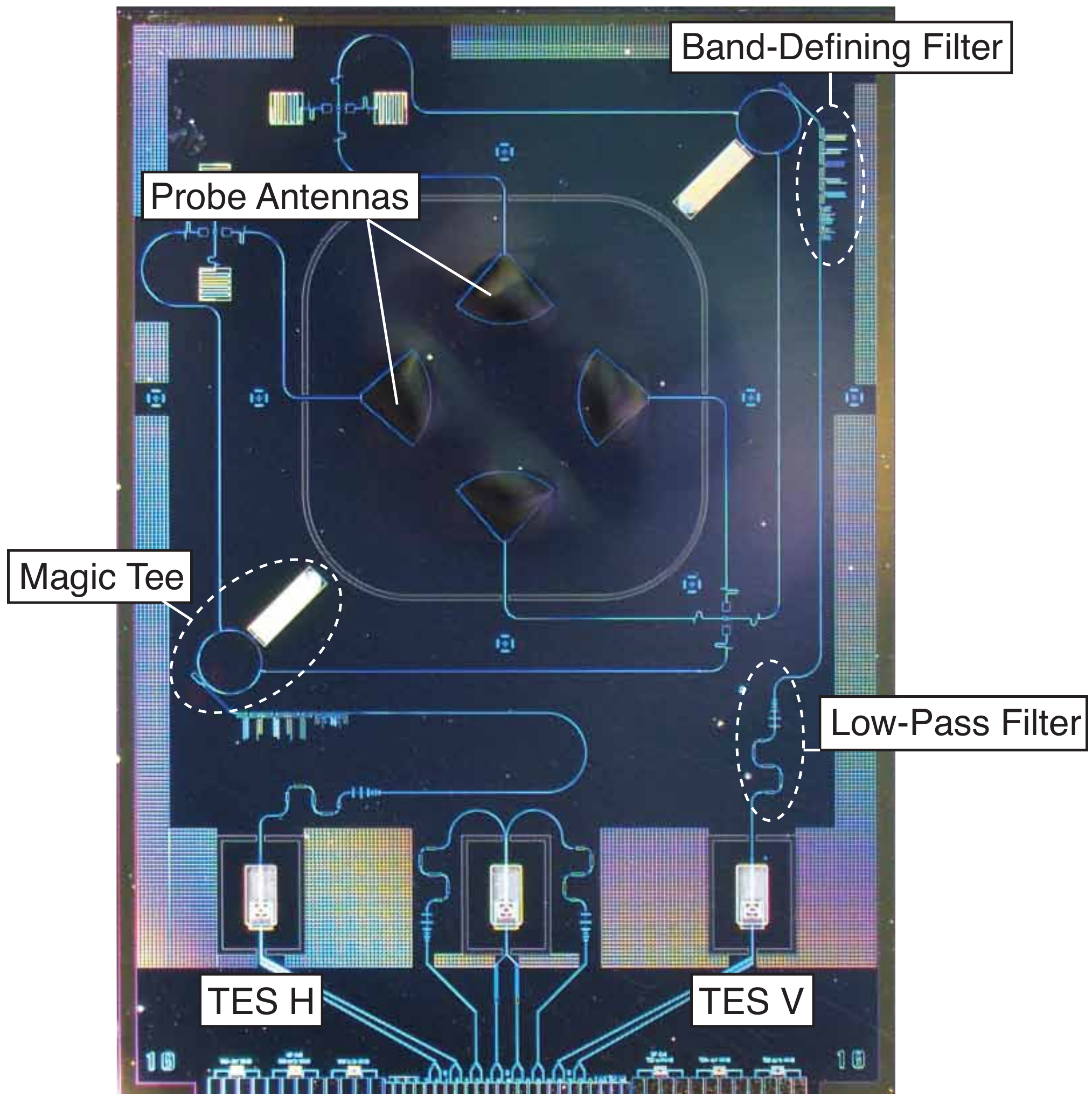}}
\subfigure[]{\includegraphics[width=7cm]{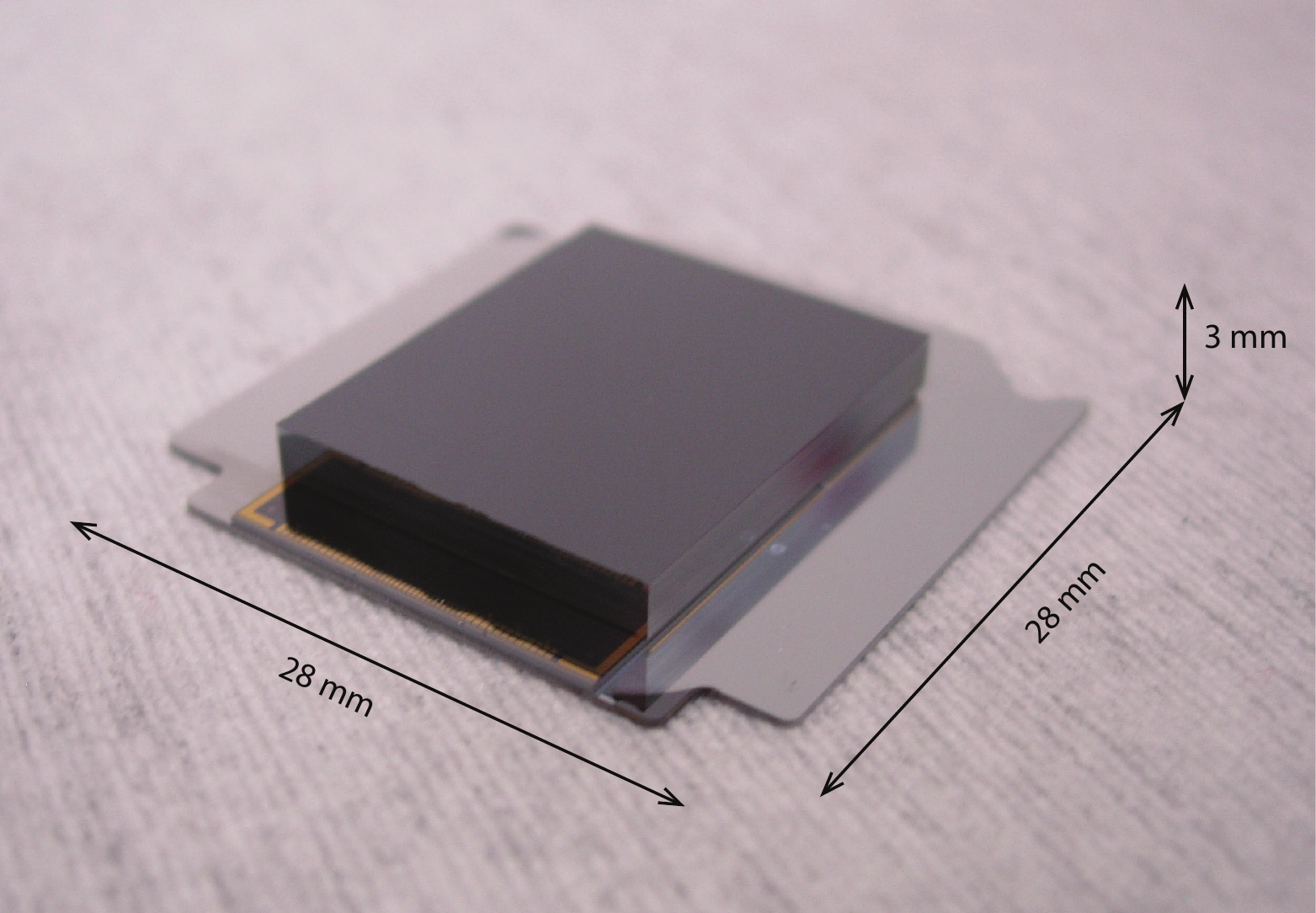}}
\caption{\label{fig:40GHz}(a) Photograph of a feedhorn-coupled Q-band detector chip. The symmetry in the microwave circuit enables broadband operation. (b) The detector chip after hybridization to the metallic enclosure, which functions as a microwave tight package that prevents high-frequency radiation from coupling to the TES bolometers. The quarter-wave backshort for the OMT probes is part of the enclosure. }
\end{center}
\end{figure}

Figure~\ref{fig:40GHz}(a) shows a CLASS Q-band detector chip. Fabrication of the detectors has been described in detail in Ref.~\citen{denis:fab}. One of the key design features of the detectors is the silicon substrate, which functions both as the dielectric for the microwave circuitry, as well as the supporting membrane for the TES bolometers and waveguide coupling probes. The single-crystal silicon is a float-zone-refined p-type boron doped layer with a room temperature resistivity of 5 k$\Omega$-cm and a carrier density of 5$\times 10^{11}$ cm$^{-3}$. 

There are several advantages to the use of single-crystal silicon. For the microwave circuitry, silicon can be micro-mashined to form a low loss dielectric substrate. An upper limit on the loss-tangent of high resistivity silicon at 4.2 K and 150 GHz~\cite{datta:siliconloss} is $\sim$7$\times$10$^{-5}$. In comparison, the loss-tangent of silicon nitride estimated from room-temperature measurements~\cite{cataldo} indicates at least a factor of 2-4 higher loss in this material~\cite{paik} when compared to silicon. From the perspective of the bolometer thermal design, the single-crystal silicon offers the advantage that the specific-heat and  thermal conductivity can be effectively modeled from theory, and reliably obtained in practice given careful design and fabrication techniques~\cite{rostem:precision}. For example, the specific heat of the silicon membrane in our devices can be estimated from the Debye equation, and is found to be small compared to the specific heat of normal metals at 0.15 K. We have successfully used Pd to set the total heat capacity of the CLASS detectors to 2.5 pJ/K with a 400 nm thick layer~\cite{rostem:precision}. Single-crystal silicon does not appear to exhibit excess heat capacity, in contrast to amorphous silicon nitride~\cite{Zink,Rostem-SiN}.   

The thermal conductance requirement for CLASS can be reliably obtained with a ballistic silicon beam 10 $\mu$m long and 13 $\mu$m wide (see Fig.~1 in Ref.~\citen{rostem:precision}, and Sec.~\ref{sec:90GHz}). A uniformity of $\pm8$\% was reported in Ref.~\citen{rostem:precision} for detectors fabricated across two wafers, and the latest results indicate a uniformity of $\pm5$\% for 36 detectors sampled across three wafers~\cite{john}. The definition of the thermal conductance with a ballistic-dominated silicon beam has now approached an engineering solution, where the required thermal conductance is obtained with a high degree of reliability given the single-crystal material, beam geometry, and appropriate fabrication technique. The remaining 5\% variability in conductance is associated with the surface roughness of the beams that support the Nb microwave and DC bias leads~\cite{rostem:precision}. The silicon surface is  roughened during dry etch steps in the fabrication chain, the statistical characteristics of which can vary significantly across a wafer and between wafers. In prototype devices where the thermal conductance was solely defined by long beams (in which phonon mean-free-path is short compared to the beam length), the non-uniformity in surface roughness caused a factor of 5 variability in thermal conductance~\cite{rostem:precision}. The effects of surface roughness on G have also been observed in silicon nitride beams~\cite{SPT90GHz}. 

We have measured the efficiency of the Q-band microwave circuitry with a cryogenic calibrator described in Ref.~\citen{rostem:calibrator}. Figure.~\ref{fig:40GHz}(b) shows the detector chip when hybridized to a metalized enclosure fabricated from a stack of silicon wafers. This enclosure mitigates the coupling of multi-moded high-frequency radiation that can potentially couple to the detectors through unintended paths in the silicon dielectric and free space. A quarter-wave backshort for the OMT probes is an element in the micro-machined silicon hybrid assembly. Preliminary measurements indicate a 90\% efficiency in-band for both channels of the Q-band polarimeter. This estimate is based on the measured band edges that are consistent with high-frequency finite-element simulations of the band-definition filter~\cite{chuss:12}. The detector response to the calibrator thermal power is linear in the 3-8 K temperature range, which suggests the microwave techniques employed for the reduction of high-frequency radiation coupling to the TES bolometers are effective to greater than 500 GHz. The microwave techniques we have employed were briefly described in Ref.~\citen{rostem:spie2012}, and will be detailed extensively in a future paper.

%%%%%%%%%%%%%%%%%%%%%%%%%%%%%%%%%%%%%%%%%%%%%%%%%%%%%%%%%%%
\section{Design of W-band Detectors}
\label{sec:90GHz}

The microwave design of the W-band detectors utilizes planar circuit elements similar to those employed at Q-band. The most notable difference between the architecture at the different frequencies is in the packaging of the detectors at the focal plane. The CLASS telescopes share the same optical design, with diffraction-limited feedhorns placed at the focal plane of each receiver~\cite{tom}. The throughput of each telescope, or $\acute{e}$tendue = $A\Omega$, is equal to $\lambda^2$ for a single mode of the optical system. Given a constant beam solid angle, the area of the feedhorn aperture, and therefore the footprint of each detector, scales with $\lambda^2$ at the focal plane. Ultimately, the feedhorn shape and size is optimized~\cite{lingzhen} for low cross-polarization, and high instrument sensitivity~\cite{tom}. For the CLASS W-band receivers, the required number of feedhorn-coupled detectors (per receiver) can be packaged into seven hexagonal monolithic arrays each with 37 dual-polarization detectors. 

\begin{figure}[!h]
\begin{center}
\subfigure[]{\includegraphics[height=7cm]{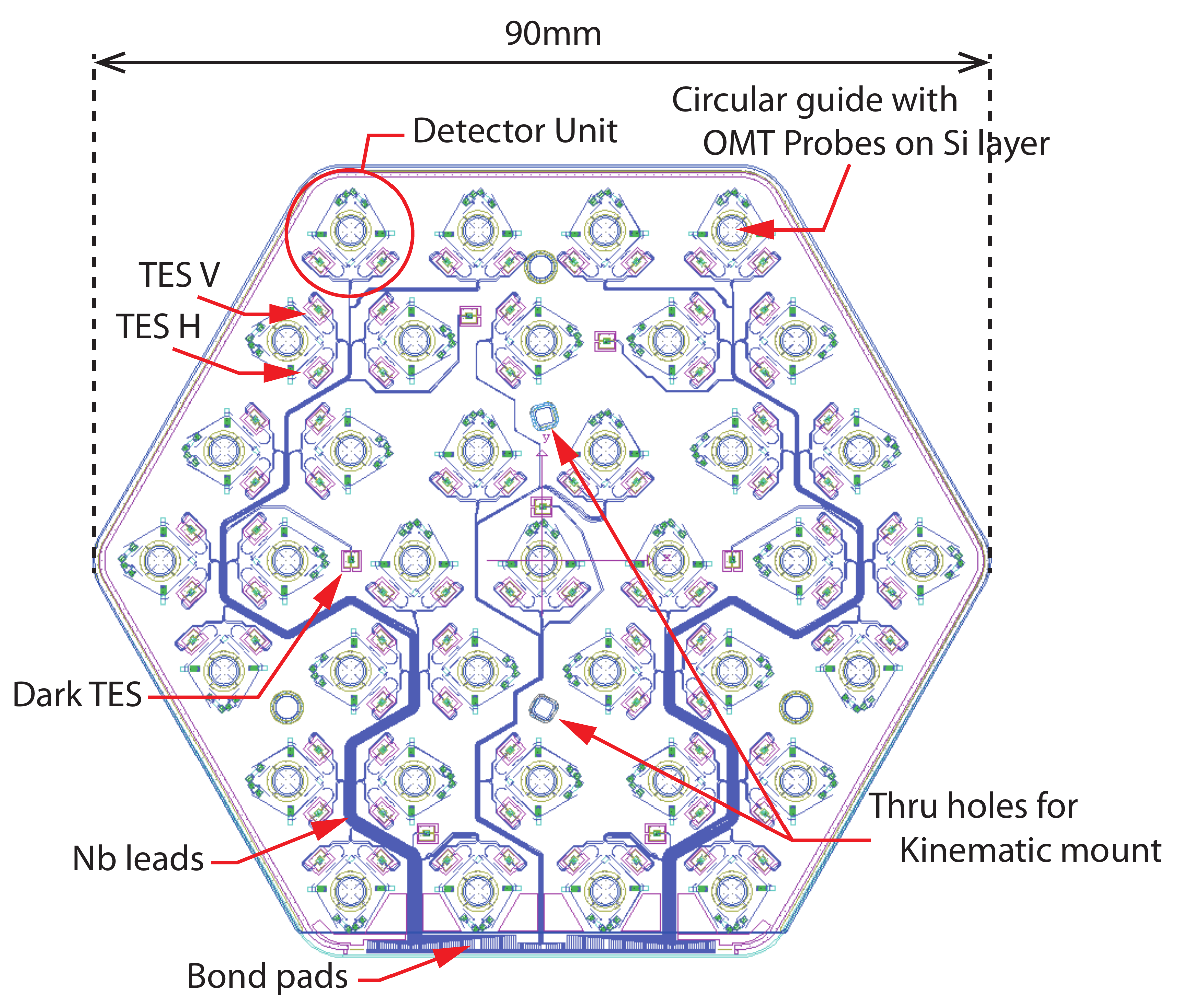}}
\subfigure[]{\includegraphics[width=7cm]{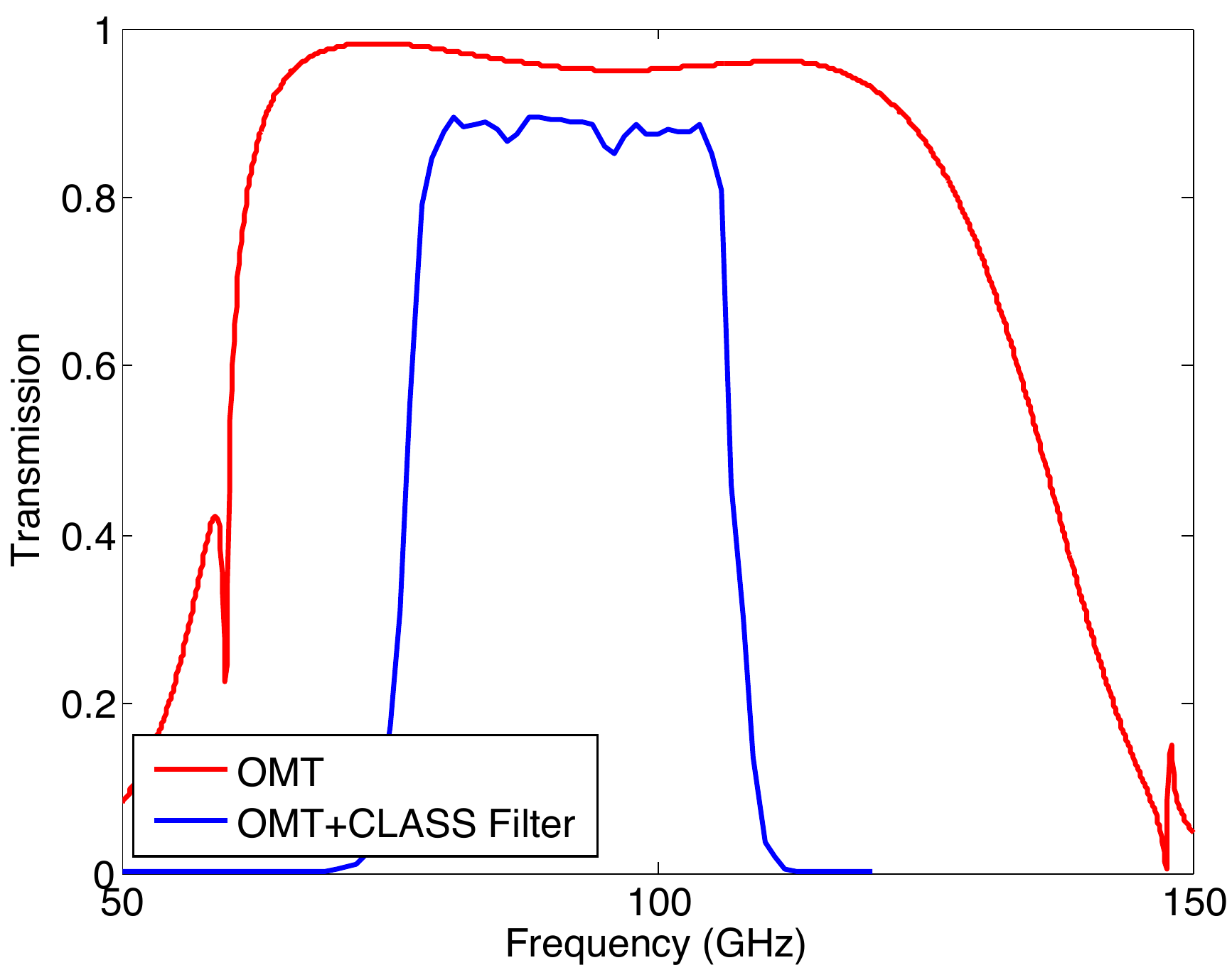}}
\caption{\label{fig:90GHzMask}(a) Design of a W-band monolithic polarimeter array. The array is fabricated from a single silicon-on-insulator wafer, and consists of 37 feedhorn-coupled detectors. (b) Full-wave simulation of an OMT (red line) and band-definition filter (blue line) response for a W-band polarimeter. The bandwidth of the reactive filter is designed to meet the requirements of the CLASS instrument~\cite{tom}.}
\end{center}
\end{figure}

Figure~\ref{fig:90GHzMask}(a) is a sketch of a W-band array fabricated on a 100 mm silicon-on-insulator wafer with a single-crystal silicon device layer 5 $\mu$m thick. The array is fabricated using the same lithographic techniques employed for the fabrication of the Q-band detectors~\cite{denis:fab}. The polarization of the sky signal is separated with an orthomode transducer (OMT), which consists of radial probes that couple each polarization to a magic-tee~\cite{yen:magicTee} (180$^\circ$ hybrid) through via-less crossovers~\cite{yen:crossover}. These components utilize the duality between microstrip and slotline~\cite{yen:scr} to realize broad bandwidth operation, low return loss, and high isolation for the appropriate modes. The OMT probes couple directly to a circular waveguide TE$^\bigcirc_{11}$ mode. The symmetry in the overall design and in each microwave component enables broadband operation over an octave from the cut-off frequency of the dominant mode. The design is flexible and enables use of the full waveguide bandwidth for space-borne or other applications as appropriate. For a sub-orbital platform, additional filters can be used to avoid atmospheric lines that can saturate the detectors~\cite{tom}. Figure~\ref{fig:90GHzMask}(b) shows the modeled response for the OMT circuit and band definition of the reactive filter employed for the W-band focal plane of the CLASS instrument. After the band definition filter, a reactive low-pass filter rejects unwanted signals from $\sim$115 GHz up to the superconducting gap frequency of Nb at $\sim$700 GHz. Above this frequency, signals are attenuated by the resistive properties of the planar transmission lines.

\begin{figure}[!t]
\begin{center}
\subfigure[]{\includegraphics[height=4cm]{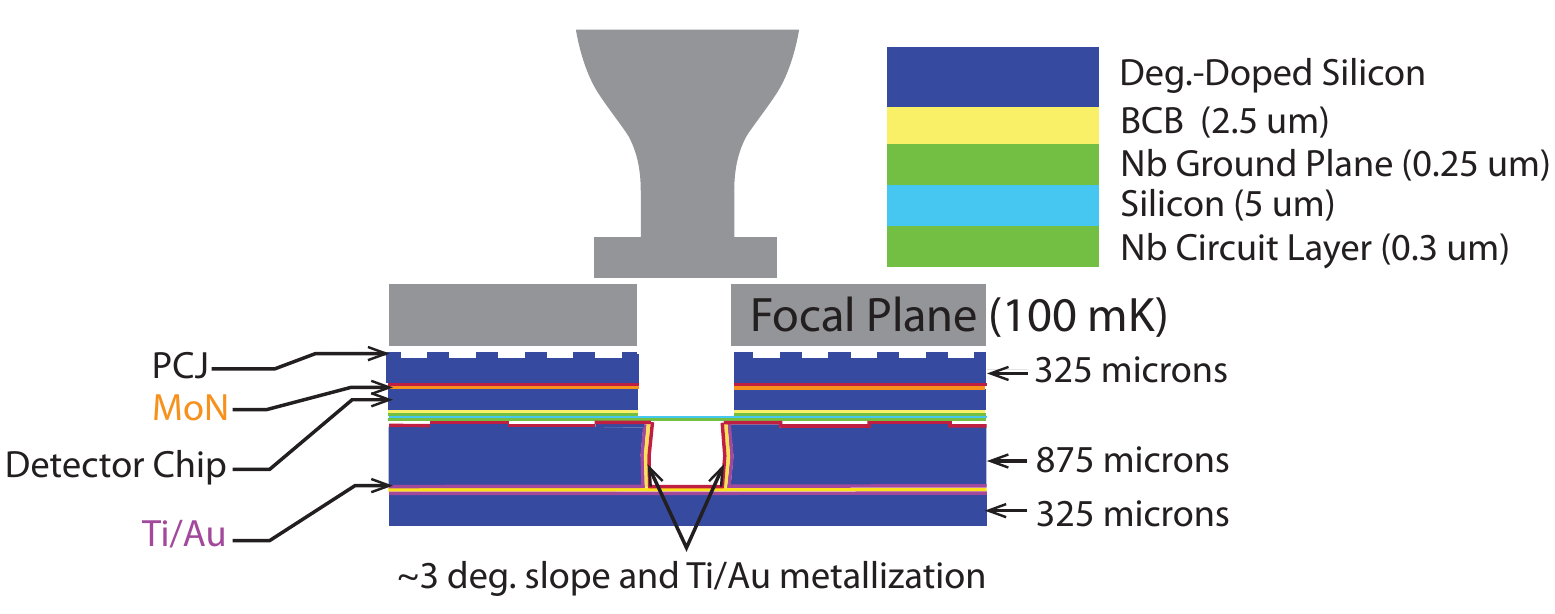}}
\subfigure[]{\includegraphics[width=6cm]{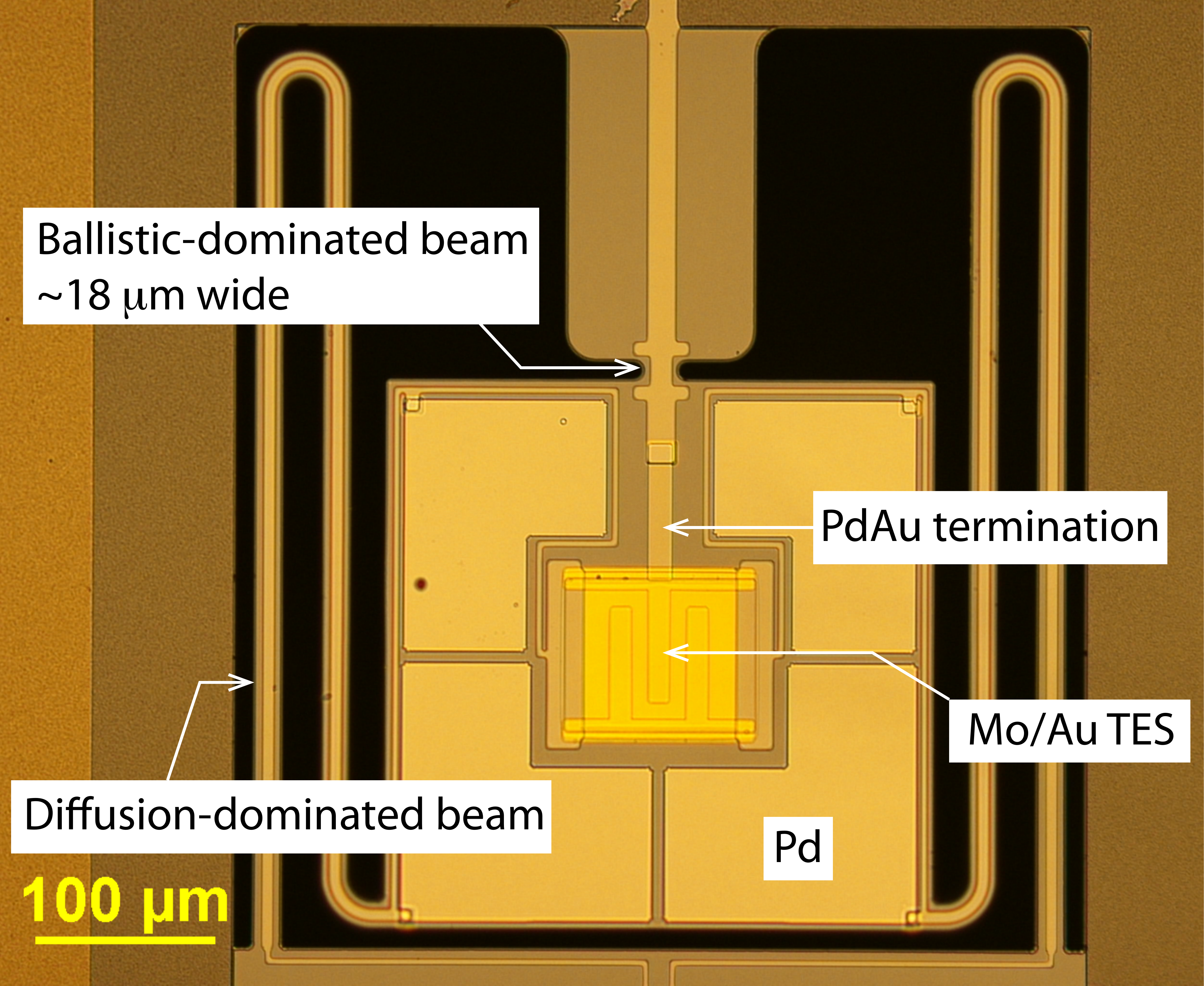}}
\caption{\label{fig:90GHzTES}(a) The W-band monolithic array is hybridized with a metalized enclosure that encapsulates the whole array. In this sketch, only one feedhorn and detector is shown in cross-section for simplicity. (b) Photograph of a W-band TES bolometer. The thermal conductance of the TES to the bath is defined by the ballistic beam. The four Pd layers define the heat capacity of the TES. The diffusion-dominated beams merely support the TES DC bias leads and have a negligible contribution to the total thermal conductance.  }
\end{center}
\end{figure}

To control radiative coupling to the bolometers, and reduce losses by reciprocity, a metalized enclosure is hybridized to the monolithic array as shown in Fig.~\ref{fig:90GHzTES}(a). The enclosure is a stack of degenerately-doped silicon wafers that includes a quarter-wave backshort for the probe antennas of each detector unit in the array. The fabrication of the enclosure is similar to that utilized for the individual Q-band detectors that has been described in Ref.~\citen{erik:fab}. The material choice and geometry of the metalized enclosure is designed to limit free-space high-frequency radiation coupling to the bolometers. For the in-band radiation, the enclosure forms a cap over the TES membranes that prevents cavity resonances. For the out-of-band multi-moded radiation, the microwave transmission and DC bias lines to the TES are boxed to form a natural waveguide high-pass filter with a cut-off frequency above $\sim1$ THz. The residual (stray) radiation coupled into the silicon dielectric is attenuated by a thin film resistive coating (5 $\Omega/\square$). Residual in-band radiation present in the sensor enclosure is absorbed by patterning this coating to achieve 150 $\Omega/\square$ structure on the dielectric substrate.

The planar transmission line design rules and fabrication process employed enable control over non-ideal circuit responses. Particular attention has gone into limiting parasitic responses in- and out-of-band. When considering surface wave losses, the substrate thickness $h/\lambda_0$ (where $h=5$ $\mu$m) is small and the loss is negligible up to 5 THz, where the first synchronous coupling of the dielectric TM$_0$ and quasi-TEM mode of microstrip occurs~\cite{Hoffmann}. In terms of radiation loss, the contribution from a slotline is proportional to the square of the slot width-to-guide wavelength ratio, and can be effectively minimized with microwave techniques that can be implemented within the micro-machining tolerances, for example, by simply reducing the slot width. Given substrate thickness and conductor widths are much smaller than the operating free-space wavelength, radiation loss from discontinuities in the circuit is a subdominant term. With these scaling laws in mind, the planar circuit design employed here can be successfully implemented at up to 300 GHz. It is worth noting that the circuit implementation at the lower frequency of Q-band is the most challenging given the greater required rejection bandwidth of out-of-band radiation. 

For the W-band monolithic array, square pillars with 4-fold rotational symmetry are lithographically machined in the silicon handle layer (Fig.~\ref{fig:90GHzTES}(a)). Once the array is in contact with, or in proximity to the focal plane baseplate, a photonic choke-joint (PCJ) is formed that creates a virtual short at the waveguide interface\cite{yen:pcj}. The PCJ allows for the thermally mismatched layers to be in contact without the need for a mechanically stiff joint. A kinematic mount\cite{tom} is used to meet the appropriate requirements on the vertical separation between the array and base plate for the proper function of the PCJ. 

%%%%%%%%%%%
For the W-band monolithic arrays, it is essential to have uniformity in detector characteristics across a wafer. Uniformity in thermal conductance $G$ and critical temperature $T_c$ of the TESs ensures that all of the detectors in an array can be biased in a single chain. The latter is not a strict requirement for an array, but simplifies readout of the detectors in an instrument with multiplexing capability. In terms of $G$, the uniformity of the Q-band detectors is excellent~\cite{john}, $\pm$5\% standard deviation at 150 pW/K. For the W-band detectors, we have reduced the number of diffusion-dominated beams by a factor two, and the ballistic-dominated beam that defines the conductance is now supporting the microwave bias line to the TES as shown in Fig.~\ref{fig:90GHzTES}(b). The length of the diffusion-dominated beams has been increased to reduce the effect from these beams on the final conductance. Preliminary measurements indicate that the total thermal conductance is 248 pW/K at 150 mK, a value that is within the predicted range for the ballistic beam~\cite{rostem:precision}. From extensive measurements of the Q-band detectors~\cite{john}, the expected percentage deviation in $T_c$ is $\pm$2\% across a wafer. Combined with the variance in $G$, the expected percentage deviation in the saturation power is $\pm$10\%, a level of uniformity that is acceptable for the multiplexing requirements. 

\section{Conclusion}

We have described the design of Q- and W-band detectors suitable for background-limited operation in space and in sub-orbital instruments. The detectors are fabricated on single-crystal silicon, which serves as a low-loss dielectric for the microwave circuitry as well as the support membrane for the Transition-Edge Sensors. The Q-band design has been successfully demonstrated. Preliminary measurements with a cryogenic blackbody load indicate $\sim$90\% efficiency in each polarization channel. Fabrication of the Q-band detectors is nearing completion, with half of the 36 polarimeters required for the CLASS instrument fabricated and tested. A important feature of the Q-band detectors is the uniformity achieved in the thermal properties of the TES bolometers. The techniques utilized to ensure this uniformity are key to the success of the W-band detectors, which are fabricated as a monolithic array of 37 detectors on a single silicon-on-insulator wafer. A micro-machined metalized enclosure is hybridized to the array at the wafer level. The enclosure controls the mode set that couples to the TES bolometers and includes the quarter-wave backshort and waveguide delay structures for the OMTÕs antenna probes.

\section{Acknowledgement}

A NASA ROSES/APRA grant provided support for the detector technology development. The CLASS project is supported by the National Science Foundation under Grant Number 0959349. T.E.-H. was supported by a National Science Foundation Astronomy and Astrophysics Postdoctoral Fellowship.

%%\bibliography{spie2014bib}   %>>>> bibliography data in report.bib

\begin{thebibliography}{10}

\bibitem{bicep2}
P.~A.~R. Ade, R.~W. Aikin, D.~Barkats, S.~J. Benton, C.~A. Bischoff, J.~J.
  Bock, J.~A. Brevik, I.~Buder, E.~Bullock, C.~D. Dowell, L.~Duband, J.~P.
  Filippini, S.~Fliescher, S.~R. Golwala, M.~Halpern, M.~Hasselfield, S.~R.
  Hildebrandt, G.~C. Hilton, V.~V. Hristov, K.~D. Irwin, K.~S. Karkare, J.~P.
  Kaufman, B.~G. Keating, S.~A. Kernasovskiy, J.~M. Kovac, C.~L. Kuo, E.~M.
  Leitch, M.~Lueker, P.~Mason, C.~B. Netterfield, H.~T. Nguyen, R.~O'Brient,
  R.~W. Ogburn, A.~Orlando, C.~Pryke, C.~D. Reintsema, S.~Richter, R.~Schwarz,
  C.~D. Sheehy, Z.~K. Staniszewski, R.~V. Sudiwala, G.~P. Teply, J.~E. Tolan,
  A.~D. Turner, A.~G. Vieregg, C.~L. Wong, and K.~W. Yoon, ``Detection of
  {$B$}-mode {P}olarization at {D}egree {A}ngular {S}cales by {BICEP2},'' {\em
  Phys. Rev. Lett.}~{\bf 112}, p.~241101, Jun 2014.

\bibitem{mortonson}
M.~J. {Mortonson} and U.~{Seljak}, ``{A joint analysis of Planck and {BICEP}2
  {B} modes including dust polarization uncertainty},'' {\em ArXiv e-prints} ,
  May 2014.

\bibitem{flauger}
R.~{Flauger}, J.~C. {Hill}, and D.~N. {Spergel}, ``{Toward an {U}nderstanding
  of {F}oreground {E}mission in the {BICEP}2 Region},'' {\em ArXiv e-prints} ,
  May 2014.

\bibitem{tom}
T.~Essinger-Hileman, A.~Ali, M.~Amiri, J.~W. Appel, D.~Araujo, C.~L. Bennett,
  F.~Boone, M.~Chan, H.-M. Cho, D.~T. Chuss, F.~Colazo, E.~Crowe, K.~L. Denis,
  R.~Dunner, J.~Eimer, D.~Gothe, M.~Halpern, K.~Harrington, G.~C. Hilton, G.~F.
  Hinshaw, C.~Huang, K.~D. Irwin, G.~E. Jones, A.~J. Kogut, D.~Larson,
  M.~Limon, L.~Lowry, N.~Mehrle, A.~Miller, N.~J. Miller, S.~H. Moseley,
  G.~Novak, C.~D. Reintsema, K.~Rostem, T.~R. Stevenson, D.~Towner, K.~U-Yen,
  E.~Wagner, D.~Watts, E.~J. Wollack, Z.~Xu, L.~Zeng, and T.~Marriage, ``The
  cosmology large angular scale surveyor ({CLASS}),'' {\em Millimeter,
  Submillimeter, and Far-Infrared Detectors and Instrumentation for Astronomy
  VII}~{\bf 9153}, p.~915353, SPIE, 2014.

\bibitem{john}
J.~W. Appel, A.~Ali, M.~Amiri, D.~Araujo, C.~L. Bennett, F.~Boone, M.~Chan,
  H.-M. Cho, D.~T. Chuss, F.~Colazo, E.~Crowe, K.~L. Denis, R.~Dunner,
  J.~Eimer, T.~Essinger-Hileman, D.~Gothe, M.~Halpern, K.~Harrington, G.~C.
  Hilton, G.~F. Hinshaw, C.~Huang, K.~D. Irwin, G.~E. Jones, J.~Krakula, A.~J.
  Kogut, D.~Larson, M.~Limon, L.~Lowry, T.~Marriage, N.~Mehrle, A.~Miller,
  N.~J. Miller, S.~H. Moseley, G.~Novak, C.~D. Reintsema, K.~Rostem, T.~R.
  Stevenson, D.~Towner, K.~U-Yen, E.~Wagner, D.~Watts, E.~J. Wollack, Z.~Xu,
  and L.~Zeng, ``The cosmology large angular scale surveyor ({CLASS}): 38-{GH}z
  detector array of bolometric polarimeters,'' {\em Millimeter, Submillimeter,
  and Far-Infrared Detectors and Instrumentation for Astronomy VII}~{\bf 9153},
  p.~915355, SPIE, 2014.

\bibitem{denis:fab}
K.~L. Denis, N.~T. Cao, D.~T. Chuss, J.~Eimer, J.~R. Hinderks, W.-T. Hsieh,
  S.~H. Moseley, T.~R. Stevenson, D.~J. Talley, K.~U.-yen, and E.~J. Wollack,
  ``Fabrication of an {A}ntenna-{C}oupled {B}olometer for {C}osmic {M}icrowave
  {B}ackground {P}olarimetry,'' {\em AIP Conference Proceedings}~{\bf 1185}(1),
  pp.~371--374, 2009.

\bibitem{datta:siliconloss}
R.~Datta, C.~D. Munson, M.~D. Niemack, J.~J. McMahon, J.~Britton, E.~J.
  Wollack, J.~Beall, M.~J. Devlin, J.~Fowler, P.~Gallardo, J.~Hubmayr,
  K.~Irwin, L.~Newburgh, J.~P. Nibarger, L.~Page, M.~A. Quijada, B.~L. Schmitt,
  S.~T. Staggs, R.~Thornton, and L.~Zhang, ``Large-aperture wide-bandwidth
  antireflection-coated silicon lenses for millimeter wavelengths,'' {\em Appl.
  Opt.}~{\bf 52}, pp.~8747--8758, Dec 2013.

\bibitem{cataldo}
G.~Cataldo, J.~A. Beall, H.-M. Cho, B.~McAndrew, M.~D. Niemack, and E.~J.
  Wollack, ``Infrared dielectric properties of low-stress silicon nitride,''
  {\em Opt. Lett.}~{\bf 37}, pp.~4200--4202, Oct 2012.

\bibitem{paik}
H.~Paik and K.~D. Osborn, ``Reducing quantum-regime dielectric loss of silicon
  nitride for superconducting quantum circuits,'' {\em Applied Physics
  Letters}~{\bf 96}(7), pp.~--, 2010.

\bibitem{rostem:precision}
K.~Rostem, D.~T. Chuss, F.~A. Colazo, E.~J. Crowe, K.~L. Denis, N.~P. Lourie,
  S.~H. Moseley, T.~R. Stevenson, and E.~J. Wollack, ``Precision control of
  thermal transport in cryogenic single-crystal silicon devices,'' {\em Journal
  of Applied Physics}~{\bf 115}(12), pp.~--, 2014.

\bibitem{Zink}
B.~Zink and F.~Hellman, ``Specific heat and thermal conductivity of low-stress
  amorphous si-n membranes,'' {\em Solid State Commun.}~{\bf 129}(3), pp.~199
  -- 204, 2004.

\bibitem{Rostem-SiN}
K.~Rostem, D.~M. Glowacka, D.~J. Goldie, and S.~Withington, ``Thermal
  conductance measurements for the development of ultra low-noise
  transition-edge sensors with a new method for measuring the noise equivalent
  power,'' in {\em Proc. SPIE},   {\bf 7020}, pp.~70200L--70200L--11, 2008.

\bibitem{SPT90GHz}
V.~Yefremenko, P.~Ade, K.~Aird, J.~Austermann, J.~Beall, D.~Becker, B.~Benson,
  L.~Bleem, J.~Britton, C.~Chang, J.~Carlstrom, H.~Cho, T.~de~Haan,
  T.~Crawford, A.~Crites, A.~Datesman, M.~Dobbs, W.~Everett, A.~Ewall-Wice,
  E.~George, N.~Halverson, N.~Harrington, J.~Henning, G.~Hilton, W.~Holzapfel,
  S.~Hoover, J.~Hubmayr, K.~Irwin, R.~Keisler, J.~Kennedy, A.~Lee, E.~Leitch,
  D.~Li, M.~Lueker, D.~Marrone, J.~McMahon, J.~Mehl, S.~Meyer, J.~Montgomery,
  T.~Montroy, T.~Natoli, J.~Nibarger, M.~Niemack, V.~Novosad, S.~Padin,
  C.~Pryke, C.~Reichardt, J.~Ruhl, B.~Saliwanchik, J.~Sayre, K.~Schafer,
  E.~Shirokoff, K.~Story, K.~Vanderlinde, J.~Vieira, G.~Wang, R.~Williamson,
  K.~Yoon, and E.~Young, ``Design and {F}abrication of 90 {GH}z {TES}
  {P}olarimeter {D}etectors for the {S}outh {P}ole {T}elescope,'' {\em IEEE
  Trans. Appl. Supercond.}~{\bf 23}(3), pp.~2100605--2100605, 2013.

\bibitem{rostem:calibrator}
K.~Rostem, D.~T. Chuss, N.~P. Lourie, G.~M. Voellmer, and E.~J. Wollack, ``A
  waveguide-coupled thermally isolated radiometric source,'' {\em Review of
  Scientific Instruments}~{\bf 84}(4), pp.~044701--6, 2013.

\bibitem{chuss:12}
D.~T. Chuss, E.~J. Wollack, R.~Henry, H.~Hui, A.~J. Juarez, M.~Krejny, S.~H.
  Moseley, and G.~Novak, ``Properties of a variable-delay polarization
  modulator,'' {\em Appl. Opt.}~{\bf 51}, pp.~197--208, Jan 2012.

\bibitem{rostem:spie2012}
K.~Rostem, C.~L. Bennett, D.~T. Chuss, N.~Costen, E.~Crowe, K.~L. Denis, J.~R.
  Eimer, N.~Lourie, T.~Essinger-Hileman, T.~A. Marriage, S.~H. Moseley, T.~R.
  Stevenson, D.~W. Towner, G.~Voellmer, E.~J. Wollack, and L.~Zeng, ``Detector
  architecture of the cosmology large angular scale surveyor,'' 2012.

\bibitem{lingzhen}
L.~Zeng, C.~Bennett, D.~T. Chuss, and E.~Wollack, ``A low cross-polarization
  smooth-walled horn with improved bandwidth,'' {\em Antennas and Propagation,
  IEEE Transactions on}~{\bf 58}, pp.~1383--1387, April 2010.

\bibitem{yen:magicTee}
K.~U-yen, E.~Wollack, J.~Papapolymerou, and J.~Laskar, ``A broadband planar
  magic-{T} using microstrip slotline transitions,'' {\em Microwave Theory and
  Techniques, IEEE Transactions on}~{\bf 56}, pp.~172 --177, jan. 2008.

\bibitem{yen:crossover}
K.~U-yen, E.~Wollack, S.~Moseley, T.~Stevenson, W.-T. Hsieh, and N.~Cao,
  ``Via-less microwave crossover using microstrip-{CPW} transitions in slotline
  propagation mode,'' in {\em Microwave Symposium Digest, 2009. MTT '09. IEEE
  MTT-S International},  pp.~1029 --1032, june 2009.

\bibitem{yen:scr}
K.~U-yen, E.~J. Wollack, S.~Horst, T.~Doiron, J.~Papapolymerou, and J.~Laskar,
  ``Slotline stepped circular rings for low-loss microstrip-to-slotline
  transitions,'' {\em Microwave and Wireless Components Letters, IEEE}~{\bf
  17}, pp.~100 --102, feb. 2007.

\bibitem{erik:fab}
E.~Crowe, C.~Bennett, D.~Chuss, K.~Denis, J.~Eimer, N.~Lourie, T.~Marriage,
  S.~Moseley, K.~Rostem, T.~Stevenson, D.~Towner, K.~U-Yen, and E.~Wollack,
  ``Fabrication of a silicon backshort assembly for waveguide-coupled
  superconducting detectors,'' {\em Applied Superconductivity, IEEE
  Transactions on}~{\bf 23}, pp.~2500505--2500505, June 2013.

\bibitem{Hoffmann}
R.~K. Hoffmann, {\em {H}andbook of {M}icrowave {I}ntegrated {C}ircuits}, Artech
  {H}ouse, 1987.

\bibitem{yen:pcj}
E.~Wollack, K.~U-yen, and D.~Chuss, ``Photonic choke-joints for
  dual-polarization waveguides,'' in {\em Microwave Symposium Digest (MTT),
  2010 IEEE MTT-S International},  p.~1, may 2010.

\end{thebibliography}
%%\bibliographystyle{spiebib}   %>>>> makes bibtex use spiebib.bst

\end{document}